\documentclass[11pt]{article}
\begin{document}

\thispagestyle{empty}

\begin{flushright} LPTENS 01/06 \end{flushright}
\vspace*{0.5cm}

\begin{center}{\bf { NON-RENORMALISATION THEOREMS IN GLOBAL SUPERSYMMETRY}}
\vskip1cm
{\bf{J. Iliopoulos}}
\vskip0.2cm

Laboratoire de Physique Th\'eorique CNRS-ENS\\
24 rue Lhomond, F-75231 Paris Cedex 05, France\\
ilio@physique.ens.fr

\vspace{0.5cm}
Invited talk at the Conference on "30 Years of Supersymmetry"

Minneapolis, Minnesota, Oct. 13-15, 2000
\vskip3.0cm
{\bf ABSTRACT}
\bigskip

We review the history of non-renormalisation theorems in global supersymmetry, 
as well as their importance in all attempts to apply supersymmetry to the real world.

\bigskip

\end{center}
\newpage

Supersymmetry in four dimensions is thirty years 
old \cite{old}, 
\cite{WZI}. It started as a rather esoteric subject, but, for the last twenty years, 
it has  occupied a unique position in elementary particle physics: It has received 
no experimental support, yet it has dominated most theoretical and much of the 
experimental work. Each one of us has his own motivations to study supersymmetry, 
but we are all fascinated by the aesthetic appeal of the theory. I have always 
considered supersymmetry as the natural extension to gauge theories. Let me explain:

We are all convinced that gauge theories have come to stay. They provide the unique 
framework,
based on deep geometrical ideas, to describe all interactions among elementary 
particles. However, they have a number of shortcomings which show that they need 
completion. The particular one I want to mention here is the fact that, to
 our present understanding, they contain three independent worlds: The world of 
radiation consists of, initially, massless vector bosons. Their number, their 
properties and their interactions are uniquely determined by the gauge group, 
they are purely geometrical objects. In all our present models the world of matter 
consists of spin one-half fermions. Their number as well as their group-theory 
properties are arbitrary, but, once assumed, they uniquely determine the interaction 
with radiation. The third world is that of Higgs scalars. They are essential for 
mass generation but they are the ones which bring most of the arbitrariness in the 
theory. In the Standard Model most free parameters are connected with the Higgs 
sector. Furthermore, their quadratic mass divergences tend to mix the various scales 
of the theory. Much effort has been devoted to constructing gauge models with Higgs 
scalars replaced by a dynamical symmetry breaking mechanism, 
but with no great success until 
now. It is therefore natural to seek a symmetry principle to relate the three worlds and obtain a trully unified theory, 
with no distinction between matter and radiation, in which all fundamental fields have
 a geometrical meaning. One is thus led to supersymmetry.

Supersymmetric field theories are "improved" field theories. The "improvement" is 
connected with the ultraviolet properties of the theory and depends on whether one 
considers theories with global or local supersymmetry. In this lecture I will review 
some of the early work on the non-renormalisation theorems in global supersymmetry. 
They are the ones which offer the possibility to solve the technical part of the gauge
 hierarchy problem.

\vskip 1 cm
The first indication that supersymmetric theories have special properties under 
renormalisation was obtained by J. Wess and B. Zumino. In their first paper on 
supersymmetry \cite{WZI} they introduced a very
simple field theory model containing a Majorana spinor, a scalar and
a pseudoscalar field. The Lagrangian density can be written as:

\begin{eqnarray}
\label{scalarmod}
{\cal{L}} & = & {\cal{L}}_0 + {\cal{L}}_m + {\cal{L}}_{int} \nonumber  \\
{\cal{L}}_0  &=  &-  {\scriptstyle {\frac{1}{2}}}(\partial A)^2 - 
{\scriptstyle {\frac{1}{2}}}(\partial B)^2 - {\scriptstyle {\frac{i}{2}}} 
\bar{\psi}\partial \!\!\!/ \psi + {\scriptstyle {\frac{1}{2}}} F^2 + 
{\scriptstyle {\frac{1}{2}}} G^2 \nonumber  \\
{\cal{L}}_m & = &  m(FA + GB - {\scriptstyle {\frac{i}{2}}}  \bar{\psi} \psi )  
\nonumber \\
{\cal{L}}_{int} & = & g(FA^2 - GB^2 + 2GAB - i \bar{\psi} \psi A + i \bar{\psi} 
\gamma_5 \psi B)
\end{eqnarray}
where $F$ and $G$ are auxiliary fields. Under the supersymmetry
transformations

\begin{eqnarray}
\label{transf}
\delta A & = & i \bar{\alpha} \psi \nonumber \\
\delta B & = & i \bar{\alpha} \gamma_5 \psi \nonumber \\
\delta \psi & = & \partial_{\mu} (A - \gamma_{5} B)  \gamma^{\mu}\alpha  + 
(F + \gamma_{5}G)\alpha    \\
\delta F & = & i \bar{\alpha} \partial \!\!\!/ \psi  \nonumber \\
\delta G & = & i \bar{\alpha} \gamma_{5} \partial \!\!\!/ \psi  \nonumber
\end{eqnarray}
\noindent
the action derived from (\ref{scalarmod}) remains invariant. In (\ref{transf}),  
$\alpha$, the parameter of the transformation, is a constant, anticommuting, Majorana 
spinor.

Upon elimination of the auxiliary fields, the model describes a
superposition of Yukawa, $\phi^{4}$ and $\phi^{3}$ couplings; it is
therefore renormalisable by power counting. Supersymmetry is manifest
by means of relations between the masses and coupling constants of
the model. Wess and Zumino tried to check by explicit calculation
whether renormalisation at the level of one loop respects these
relations \cite{WZII}. To their surprise, they discovered that it did much more:
The only necessary counterterm was a single wave function
renormalisation common to all fields. Neither mass nor coupling
constant renormalisations were needed.

I was very sceptical when I first heard there results. I was inclined
to believe that they were accidents of one loop and they would not
survive at higher orders. {\it Par acquit de conscience} we decided
with B. Zumino to check the two loop level. You guess the answer:
The same result holds, although, this time the cancellation involves
diagrammes with different topologies. It was clear that a general proof should exist.
Indeed, it turned out to be rather simple \cite {IZ}.

As I said before, the model is renormalisable by power counting. Furthermore, because 
of the fact that the transformations (\ref{transf}) are global ($\alpha$ constant), 
one can find easily supersymmetry preserving regularisation schemes. 
For example, a higher derivative kinetic term will do \cite {IZ}. It follows that 
renormalisation will preserve supersymmetry, in other words one will need, at most, 
three counterterms, $Z$, ${\delta}m$ and $Z_g$, one for each term in (\ref{scalarmod}). The surprising result was that only $Z$ was needed. These extra cancellations that give  ${\delta}m = 0$ and $Z_g = 1$ clearly go beyond symmetry alone since the corresponding terms ${\cal{L}}_m$ and  ${\cal{L}}_{int}$ are allowed by supersymmetry.

The particular property of (\ref{scalarmod}) which is relevant for the proof is 
\cite {IZ}:

\begin{equation}
\label{prop}
\frac{\partial}{{\partial}m}{\cal{L}}_m = \frac{1}{2g}\frac{\partial}{\partial A}
{\cal{L}}_{int}
\end{equation}

This property has an immediate analogue in terms of the 1P-I functions: Let 
$\Gamma [R]$ be the generating functional where  $R$ denotes, collectively, the 
classical fields, the conjugate variables under Legendre transformation of the 
external sources. We obtain:

\begin{equation}
\label{prop1}
\frac{\partial}{{\partial}m} \Gamma [R] = -\frac{m}{2g} \int R_F(y)d^4 y + 
\frac{1}{2g} \int \frac{\delta \Gamma [R]}{\delta R_A (y)} d^4 y
\end{equation}
\noindent
which means that, for 
every vertex function other than the 
1P-I part of $<F>_0$, the derivative with respect to the bare mass 
gives an insertion of a zero momentum $A$ field. On the other hand, using the 
supersymmetry Ward identities and the equations of motion of the regularized theory, 
it is easy to prove \cite {IZ} that the vacuum expectation values of all fields 
vanish. We now take the functional derivative of (\ref{prop1}) with respect to 
$R_F$ and then put all $R$'s equal to zero. Using the vanishing of $<F>_0$ we 
obtain:

\begin{equation}
\label{prop2}
m = Z^{-1} \Gamma _{FA} (p^2 = 0)
\end{equation}
\noindent
which implies

\begin{equation}
\label{prop3}
m_r = Zm
\end{equation}
\noindent
i.e. ${\delta}m = 0$. Similarly we obtain

\begin{equation}
\label{prop4}
g_r = Z^{\frac{3}{2}} g
\end{equation}
\noindent
i.e. $Z_g = 1$. This completes the proof of the non-renormalisation theorem.

In terms of the renormalisation group functions (\ref{prop4}) has an interesting 
consequence \cite{FIZ}:

\begin{equation}
\label{rengr}
\beta (g) = 3 g \gamma (g)
\end{equation}

Using (\ref{rengr}), it is easy to show that the $\beta$ function of this model 
cannot have a non-trivial fixed point. Indeed, let us consider the massless case. 
A fixed point $g_0$ satisfies $\beta(g_0) = 0$. 
But then (\ref{rengr}) implies that $\gamma(g_0)$ also vanishes, i.e. all fields have 
canonical dimensions and all Green functions satisfy free-field theory 
renormalisation group equations. This is enough to show that the model is in fact 
free. But $g$ is defined as the value of the three point function at zero external 
momenta and cannot be non-zero for a free field theory. Turning to the massive case 
we can write the corresponding Callan-Symanzik equation:

\begin{equation}
\label{CS}
[m  \frac{\partial}{\partial m} + \beta (g) - n \gamma (g) ] 
{\Gamma}_{\phi_1 ... \phi_n}(p_i ;m,g) = \frac{m}{2g} \delta (g) 
{\Gamma}_{A, \phi_1 ... \phi_n}(0, p_i ;m,g)
\end{equation}
\noindent
where $m$ and $g$ denote the renormalised quantities and

\begin{equation}
\label{CS1}
\beta (g) = {\scriptstyle {\frac{3}{2}}} g \frac{f}{1+f} \qquad , 
\qquad  \gamma (g) = {\scriptstyle {\frac{1}{2}}} \frac{f}{1+f} \qquad , 
\qquad \delta (g) = \frac{1}{1+f}
\end{equation}

Equation (\ref{CS}) has some interesting features \cite{FIZ} : First, all three 
functions usually appearing, namaly $\beta$, $\gamma$ and $\delta$, are expressed 
in terms of the single function $f$. This is a direct consequence of the 
non-renormalisation theorem. In perturbation $f$ has a power series expansion in $g$:

\begin{equation}
\label{CS2}
f(g) = \frac{g^2}{4 \pi^2} +...
\end{equation}

Second, at the right hand side of (\ref{CS}), instead of the familiar mass insertion, 
there appears an insertion of a zero momentum $A$ field. Notice that, since the 
added line carries zero momentum, the Green function in the left hand side still 
dominates in the deep Euclidean region.

At this stage these divergence cancellations appeared to be miraculous.
We had no deeper understanding of their origin and we could only
speculate. The first question was whether the remaining divergence, the
wave function counterterm, was really present to all orders, in other
words the question was whether the theory was in fact
superrenormalisable. Our explicit calculation showed that, at least
up to two loops, this did not seem to be the case, although a partial
cancellation did occur \cite{IZ}. A related question was whether
supersymmetry could turn a  theory which is non-renormalisable by
power counting into a renormalisable one. We could find no examples.
Today we know the precise answer to such questions. We use a new
formulation of supersymmetry \cite{supspace}, in which
the base space is eight dimensional. Four are the usual Minkowski
coordinates $x_{\mu}$, $\mu = 0,1,2,3$ and the remaining four can be
viewed as forming, under Lorentz transformations,  a complex
two-component Weyl spinor
${\theta}_{\alpha}$ together with its conjugate $\bar{\theta}_{\dot{\alpha}}$,
$\alpha , \dot{\alpha } = 1,2$. These components are taken to be totally
anti-commuting
elements of a Grassmann algebra. This space is called "superspace".
We can show that supersymmetry transformations act on superspace as
generalised translations.
The important point is that, because of the anti-commutation properties
of the components of $\theta$, any function in superspace is, in fact,
a polynomial in $\theta$ and $\bar{\theta}$ with coefficients which are
functions of $x$. In other words, a field in superspace, called
"superfield", is equivalent to a finite multiplet of ordinary fields.

\begin{equation}
\label{supfield}
\Phi (x, \theta , \bar{\theta} ) = A(x)+\theta \psi (x) + \bar{\theta}
\bar{\chi} +.....+\theta \theta \bar{\theta} \bar{\theta} R(x)
\end{equation}

The fields $A(x)$, $\psi (x)$, etc which appear as coefficients in
the expansion (\ref{supfield}), have well-defined Lorentz transformation
properties and transform among themselves and their derivatives under
supersymmetry. Therefore, they form a representation, in general
reducible. It is possible to find a complete set of covariant restrictions
on superfields to obtain the irreducible representations. A particular
example of such restrictions is given by the equation:

\begin{equation}
\label{chiral}
\frac{\partial}{\bar{\theta}_{\dot{\alpha}}}\Phi (x, \theta ,\bar{\theta}) = 0
\end{equation}

A superfield that satisfies (\ref{chiral}) is a function of $x$ and
$\theta$ only and is called "chiral". Its expansion in powers
of $\theta$ is given by:

\begin{equation}
\label{chiral1}
\phi (x, \theta ) = A(x) + \theta \psi (x) + \theta \theta F(x)
\end{equation}

It is precisely the multiplet we considered in (\ref{scalarmod}) written in 
complex notation.

These
observations provide the basis for the representation theory of
supersymmetry. Since the product of two superfields is again a
superfield, they also provide the elements of a tensor calculus. The
Lagrangian densities which were initially constructed by trial and
error can be obtained now as superfields in superspace. The
corresponding actions are eight dimensional integrals of the Lagrangian
superfields. Integrals over Grassmann variables have unusual properties. In 
particular, they satisfy:

\begin{equation}
\label{Grasint}
\int d\theta = 0  \qquad , \qquad \int \theta  d\theta = 1
\end{equation}
\noindent
which means that only the last term in the expansion of a superfield survives. 
With these results we can reformulate perturbation theory using Feynman rules 
directly in superspace \cite{Gris..}. It follows that, although we can 
use a chiral superfield as part of a Lagrangian and integrate this piece only over 
$\theta$ to obtain the action, only integrals over the entire superspace appear as 
counterterms. Thus we can understand the origin of the non-renormalisation theorems. 
Going back to (\ref{scalarmod}), we can show that both ${\cal L}_m$ and 
${\cal L}_{int}$ are chiral superfields and depend only on $\theta$, while 
${\cal L}_0$ depends on both $\theta$ and $\bar{\theta}$. Therefore only a wave 
function counterterm will appear.

Before leaving the scalar model I want to address the following question: After all, 
the supersymmetric model (\ref{scalarmod}) is just a particular combination of Yukawa 
and scalar couplings. Are there any other combinations with similar remarkable 
properties? As an example, let us consider the massless Lagrangian \cite{INT}:

\begin{eqnarray}
\label{genmod}
{\cal{L}} = & - &{\scriptstyle {\frac{1}{2}}}(\partial A)^2 - 
{\scriptstyle {\frac{1}{2}}}(\partial B)^2 - {\scriptstyle {\frac{i}{2}}} 
\bar{\psi}\partial \!\!\!/ \psi   \nonumber  \\
& - & ig \bar{\psi} \psi A + i g \bar{\psi} \gamma_5 \psi B  \\
& - &{\scriptstyle {\frac{1}{2}}} \lambda (A^4 + B^4 ) - fA^2 B^2 \nonumber
\end{eqnarray}

Invariance under supersymmetry is obtained for 

\begin{equation}
\label{relat}
\lambda = f = g^2
\end{equation}

 At one loop we can 
compute the counterterms of this model \cite{INT} with the following results:
(i)Only under the supersymmetric relation (\ref{relat}) we obtain the 
non-renormalisation theorem. (ii)As expected, this relation is a fixed point 
of the renormalisation group flow. (iii)This fixed point is an infrared attractor,
i.e. if we start at high energies somewhere in its vicinity, we shall be driven
towards it at larger and larger distances.

The superfield techniques have been used to derive more general non-renormalisation
theorems and they will be reviewed elsewhere. In the rest of my time I want to
describe some early results on supersymmetry breaking.

In Nature we see no degeneracy between fermions and bosons. So it was immediately recognised that supersymmetry, if at all relevant, must be broken. In our first paper with Zumino \cite{IZ} we addressed the question for the simple model of equation (\ref{scalarmod}). An ordinary internal symmetry is spontaneously broken if an operator, usually one of the canonical fields, which transforms non trivially under the symmetry transformations, is allowed to take a non-zero vacuum expectation value. This is often  achieved by choosing a negative value for the mass square term of a scalar field.  In supersymmetry the situation turns out to be different. The masses of scalars and spinors are digenerate and the classical potential for any scalar field never becomes negative. In our model the only fields which can take non-vanishing vacuum expectation values without breaking Lorentz invariance or parity are $A$ and $F$. However the first one does not help because $A$ appears only in th!
e !
transformation law of $\psi$ thr
ough its derivative (\ref{transf}). Therefore, as long as we do not break translational invariance, we can shift it with a constant value without breaking supersymmetry. We are left with the $F$ field. It transforms by a total derivative (\ref{transf}) and therefore,  we can add to the Lagrangian (\ref{scalarmod}) a term linear in $F$ without breaking supersymmetry explicitly. But even this term does not help because it is straightforward to verify that it can always be eliminated by a shift of the $A$ field. We thus showed that there was no spontaneous breaking of supersymmetry, for any choice of the parameters of the model. We also asked the general question of the  possibility of spontaneous breaking and we gave the wrong answer. We argued that, since the hamiltonian of a supersymmetric system can be expressed as the anti-commutator of two fermionic generators, the energy of an eigenstate which is annihilated by the generators is zero. On the other hand, the same relation!
 s!
hows that the spectrum of the ha
miltonian is positive semi-definite. Therefore we concluded that the supersymmetric invariant state will be always the ground state. We were aware of the fact that the statement was not rigorous and we said so, but we believed the result to be correct. I shall come back to the fate of spontaneous breaking in a moment, but let me examine first the consequences of a soft, explicit breaking for the model of equation (\ref{scalarmod}).

The simplest breaking term is one linear in the field $A$ \cite{IZ}:

\begin{equation}
\label{break}
{\cal{L}}  \to  {\cal{L}} -cA
\end{equation}

This term can be eliminated by a simultaneous shift of the fields $A$ and $F$, 
$A \rightarrow A+a$, $F \rightarrow F+f$ with $a$ and $f$ constants given in terms of the original parameters $m$, $g$ and $c$. The shift in $F$ breaks supersymmetry and induces a mass-splitting in the multiplet:
\newpage

\begin{eqnarray}
\label{massform}
m_{\psi} & = & m +2ga = c/f \nonumber \\
{m_A}^2 & = & {m_{\psi}}^2 - 2fg \\
{m_B}^2 & = & {m_{\psi}}^2 + 2fg \nonumber
\end{eqnarray}

The masses are no longer equal but, in the tree approximation, they satisfy the relation:

\begin{equation}
\label{massrel}
{m_A}^2 +{m_B}^2 = 2 {m_{\psi}}^2
\end{equation}

The remnant of the non-renormalisation theorem presented above, guarantees that this relation will receive no divergent corrections in higher orders.

If one eliminates $f$, one finds an equation of third degree for $a$. For small but finite $c$, its solutions correspond to the extrema of the potential for the field $A$:

\begin{equation} 
\label{potent}
V(a) =  {\scriptstyle {\frac{1}{2}}}a^2 (m+ga)^2 + ca
\end{equation}

As $c \rightarrow 0$, the three solutions become

\begin{equation} 
\label{solut}
a_1 =0  \qquad , \qquad a_2 = -\frac{m}{g}  \qquad , \qquad a_3 =  -\frac{m}{2g}
\end{equation}
\noindent
and the potential becomes symmetric around the value $a_3$. $a_1$ and $a_2$ correspond to the two minima of the potential and give two, stable, supersymmetric, physically equivalent solutions. $a_3$ corresponds to a local maximum and it is unstable. It is instructive to notice that if one could choose this unstable solution, one would have $m_{\psi} = 0$, i.e. the $\psi$ field would become a Goldstone spinor and supersymmetry would be spontaneously broken \cite{IZ}. In this case the relation (\ref{massrel}) shows that one of the bosons would have a negative square mass. This is the sign of instability.

This mass relation which shows an equal splitting among the levels in broken supersymmetry turns out to be very general in the breaking of both global ant local supersymmetries \cite{FGP}. In fact it poses severe constraints in model building. Although, in general, quantum corrections are expected to modify it, it turns out that it is remarkably robust \cite{GI}.

I shall end with a short review of the mechanism of spontaneous  breaking of global supersymmetry. As I explained before, the particular connection between supersymmetry generators and translations, shows that a supersymmetric invariant state is always a ground state \cite{IZ}. This was initially interpreted as an indication for the existence of a no-go theorem as regards to spontaneous supersymmetry breaking. In fact the situation is different: It is correct that a supersymmetric invariant state, if it exists,  is always a ground state and global supersymmetry is unbroken. But, contrary to what happens in ordinary symmetries, a supersymmetric state may not exist at all. In this case, and in this case only, spontaneous breaking occurs. The first example \cite{FI} was that of the supersymmetric extension of a $U(1)$ gauge theory \cite{WZIII}. The model describes the interaction of a charged scalar multiplet and a gauge vector multiplet. In a particular family of gauge choices!
, !
the Wess-Zumino gauge \cite{WZII
I}, the Lagrangian is polynomial and renormalisable by power counting. In terms of physical fields, the gauge multiplet consists of the photon field ${\cal{V}_{\mu}}(x)$ and its supesymmetric partner which is a neutral Majorana spinor $\lambda (x)$. The matter multiplet consists of a Dirac spinor $\psi (x)$ (the electron) and two charged spin zero fields, a scalar $A(x)$ and a pseudoscalar $B(x)$. The photon has the usual electromagnetic couplings with the charged fields, characterised by a coupling constant $e$ and  supersymmetry induces new, Yukawa type couplings between $\lambda$, $\psi$ and $A$ and $B$,  with a coupling constant which is again equal to $e$. As with eq.(\ref{scalarmod}), the transformations are simpler if one includes auxiliary fields, a charged scalar $F$, a charged pseudoscalar $G$ and a neutral  pseudoscalar $D$. The first two are associated with the matter multiplet and the last one with the photon. Under supersymmetry transformations $D$ has properti!
es!
 similar to those of $F$ and $G$
 of eq.(\ref{transf}), i.e. it transforms by a four derivative and it appears without derivative in the transformation of $\lambda$. Therefore, we can add to the Lagrangian a term linear in the field $D$. This term preserves  supersymmetry and gauge invariance and violates parity explicitly but softly. On the other hand, a non zero vacuum expectation value for $D$ breaks supersymmetry spontaneously. The classical potential for the spin zero fields is given by:

\begin{eqnarray}
\label{qed}
V & = & {\scriptstyle {\frac{1}{2}}}[{F_1}^2 + {F_2}^2 + {G_1}^2 + {G_2}^2 + D^2] \nonumber \\
& + & m(F_1 A_1 +F_2 A_2 + G_1 B_1 + G_2 B_2) \\
& + & e D(A_1 B_2 -A_2 B_1) +\xi D \nonumber 
\end{eqnarray}

\noindent
where $F_1$, $F_2$ etc are the real and imaginary parts of the fields. The novel feature here is that the linear term $\xi D$ cannot be absorbed by a shift of an $A$ or $B$ field and, therefore, we expect supersymmetry to be spontaneously broken \cite{FI}. Indeed, after elimination of the auxiliary fields and diagonalisation of the resulting mass terms, we obtain:

\begin{equation}
\label{spbrmass}
{\cal{L}}_m =- {\scriptstyle {\frac{1}{2}}}(m^2 + e\xi)({{\tilde{A}}_1}^2 + {{\tilde{B}}_1}^2) - {\scriptstyle {\frac{1}{2}}}(m^2 - e\xi)({{\tilde{A}}_2}^2 + {{\tilde{B}}_2}^2) - i m \bar{\psi} \psi
\end{equation}

\noindent
where ${\tilde{A}}_i$ etc are linear combinations of the old fields. ${\cal{V}_{\mu}}$ and $\lambda $ remain massless. In fact it
is easy to show that 
$\lambda$ is the Goldstone spinor one expects after spontaneous
breaking of a symmetry whose conserved current has spin equal to
$3/2$. We can verify this result explicitly by studying the
transformation properties of the fields under infinitesimal
supersymmetry transformations. A Goldstone field is the one which has
in its transformation law a constant term which is not proportional to
any other field. 

The relations (\ref{spbrmass}) show that we can distinguish two cases
depending on the sign of $ m^2 - e\xi $. (We assume, without loss of
generality, $e\xi >0 $). The positive sign means that supersymmetry is
spontaneously broken but gauge symmetry is not. In the oposite case
they are both spontaneously broken and the photon becomes massive by
the usual Brout-Englert-Higgs mechanism. The Goldstone fermion is now
a linear combination of $\lambda$ and $\psi$. 

This simple example shows the general mechanism for spontaneous
supersymmetry breaking. In fact, it would have been impossible to have
such a breaking, if it were not for the peculiar property we mentioned
earlier, namely the possibility of adding to the Lagrangian a term
linear in the auxiliary fields without breaking supersymmetry
explicitly. If we restrict ourselves to renormalisable
theories, we can use only scalar and vector multiplets with auxiliary
fields we have called brfore $F$, $G$, and $D$. The first is scalar,
the other two pseudoscalar. Let $\phi$ denote, collectively, all other
physical, spin zero fields. We shall assume that Lorentz invariance is
not broken, consequently all other fields have zero vacuum expectation
values. The potential of the scalar fields in the tree approximation
has the form:

\begin{eqnarray}
\label{scpot}
V(\phi)=-{\scriptstyle {\frac{1}{2}}}[\Sigma {F_i}^2 +\Sigma
{G_i}^2 +\Sigma {D_i}^2  ]\nonumber  \\  +  [\Sigma F_i F_i (\phi)
+\Sigma G_i G_i (\phi) +\Sigma D_i D_i (\phi)  ]    
\end{eqnarray}

\noindent
where the functions $F_i(\phi)$, $G_i(\phi)$ and $D_i(\phi)$ are
polynomials in the physical fields $\phi$ of degree not higher than
second. The equations of motion which eliminate the auxiliary fields
are

\begin{equation}
\label{auxfields}
F_i=F_i(\phi) \ ; \ G_i=G_i(\phi) \ ; \ D_i=D_i(\phi) 
\end{equation}
 
\noindent
so the potential in terms of the physical fields reads:

\begin{equation}
\label{phypot}
V(\phi)={\scriptstyle {\frac{1}{2}}}[\Sigma {F_i}^2 (\phi)+\Sigma
{G_i}^2 (\phi)+\Sigma {D_i}^2 (\phi)] 
\end{equation}

The important point is that $V$ is non-negative and vanishes only when

\begin{equation}
\label{eqspbr}
F_i(\phi)=0 \ ; \ G_i(\phi)=0 \ ; \ D_i(\phi)=0 
\end{equation}

\noindent
i.e. when all auxiliary fields have zero
vacuum expectation values and supersymmetry is unbroken. For
spontaneous breaking we must arrange so that the system  of the
second degree algebraic equations (\ref{eqspbr}) has no real
solution. This was the case in the supersymmetric extension of
Q.E.D. we presented before. We can also construct models with more
than one scalar multiplets \cite{FO}. A final remark: The
non-renormalisation theorems we presented before show that, if global 
supersymmetry is unbroken in the tree approximation, it will remain
unbroken to all orders in perturbation theory. This also puts severe
restrictions in model building where non-perturbative breaking
mechanisms must be invented \cite{W}.  

As we saw in the previous example and as we know from general theorems,
spontaneous breaking of supersymmetry results in the appearance of a
zero mass Goldstone spinor. It satisfies the standard low-energy
theorem, known as ``Adler's zero''. It states that the amplitude for
the emission (or absorption) of a Goldstone particle of momentum $k$
vanishes at low energies linear in $k$. This means that this fermion
cannot be identified with one of the neutrinos of the Standard Model,
even if they have exactly zero mass \cite{WF}. Although we have no experimental
hint of any kind, the predominant philosophy to-day is to believe that
such a Goldstone fermion is absorbed in a super-Higgs mechanism in the
framework of a supergravity theory \cite{VS}.

It is still too early to take bets on the final place that
supersymmetry will occupy in particle physics. It is amusing  to
notice that several times in recent years, whenever experimental results 
appeared to depart from the Standard Model predictions, the first
reaction of both theorists and experimentalists  was to try to 
interpret them as indications of supersymmetry. Few theories have
exercised so much fascination to so many physicists for so long. Looking for supersymmetric
particles will be an important part of experimental research in the
years to come. I hope that it will be both exciting and rewarding and
that the fortieth anniversary Conference will be that of
supersymmetric phenomenology.

\end{document}